\documentclass[10pt,twocolumn]{ICCAS}
 
\usepackage{diagbox}

\usepackage{siunitx} 
\usepackage[version=4]{mhchem} 
\usepackage{cleveref}
\usepackage{tikz}
\usetikzlibrary{arrows.meta,positioning}
\usepackage{tabularx}
\usepackage{csquotes}
\usepackage{url}
\usepackage[abbreviations,acronym,prefix,shortcuts=ac,section=subsection,toc=false,symbols]{glossaries-extra}

\setabbreviationstyle[\acronymtype]{long-short}

\glsdisablehyper 

\renewcommand{\glossarysection}[2][]{} 

\newacronym[
longplural={battery electric buses},
]{BEB}{BEB}{battery electric bus}
\newacronym[prefixfirst={a\ }, 
prefix={an\ }, 
]{EV}{EV}{electric vehicle}
\newacronym[
prefixfirst={a\ }, 
prefix={an\ }, 
]{SOC}{SOC}{state of charge}
\newacronym{OCP}{OCP}{optimal control problem}
\newacronym[
prefixfirst={a\ }, 
prefix={an\ }, 
]{MPC}{MPC}{model predictive control}
\newacronym{DC}{DC}{direct current}
\newacronym{OCV}{OCV}{open-circuit voltage}
\newacronym{ODE}{ODE}{ordinary differential equation}
\newacronym{BESS}{BESS}{battery energy storage system}
\newacronym{SOE}{SOE}{state of energy}
\newacronym{V2G}{V2G}{vehicle-to-grid}
\newacronym{BMS}{BMS}{battery management system}
\newacronym{PV}{PV}{photovoltaics}
\newacronym{EMS}{EMS}{energy management system}
\newacronym[
prefixfirst={a\ }, 
prefix={an\ }, 
]{NLP}{NLP}{non-linear program}
\newacronym[
longplural={degrees of freedom},
]{DOF}{DOF}{degree of freedom}
\newacronym{GTFS}{GTFS}{General Transit Feed Specification}
\newacronym{TOU}{TOU}{time-of-use}
\newcommand{\Pchgbus}{\ensuremath{P_{\mathrm{chg},u}}}
\newcommand{\PchgbusLim}{\ensuremath{\overline{P}_{\mathrm{chg},u}}}

\newcommand{\Pibus}{\ensuremath{P_{\mathrm{chg,i},u}}}

\newcommand{\Pdemandbus}{\ensuremath{P_{\mathrm{dem},u}}}
\newcommand{\PdemandbusRealization}{\ensuremath{\widetilde{P}_{\mathrm{dem},u}}}
\newcommand{\PdemandbusWorst}{\ensuremath{P_{\mathrm{dem},u}^{\mathrm{wc}}}}
\newcommand{\Pgrid}{\ensuremath{P_{\mathrm{grid}}}}
\newcommand{\PgridUpperLim}{\ensuremath{\overline{P}_{\mathrm{grid}}}}

\newcommand{\Pmax}{\ensuremath{P_{\mathrm{grid}}^{\mathrm{max}}}}
\newcommand{\bus}{\ensuremath{u}}
\newcommand{\busSet}{\ensuremath{{U}}}
\newcommand{\Nbus}{\ensuremath{M}}
\newcommand{\timeVar}{\ensuremath{t}}
\newcommand{\tInit}{\ensuremath{t_{0}}}
\newcommand{\tFinal}{\ensuremath{t_{\mathrm{F}}}}
\newcommand{\timeStep}{\ensuremath{\Delta{}t}}
\newcommand{\Ebus}{\ensuremath{E_{u}}}
\newcommand{\EbusWorst}{\ensuremath{E_{u}^{\mathrm{wc}}}}

\newcommand{\EbusLim}{\ensuremath{\overline{E}_{u}}}

\newcommand{\effbus}{\ensuremath{\eta_{\mathrm{u}}}}

\newcommand{\chgLossBus}{\ensuremath{\alpha_{\mathrm{u}}}}

\newcommand{\inDepot}{\ensuremath{\delta_{u}}}
\newcommand{\demandUncertainty}{\ensuremath{\rho}}
\newcommand{\specificEnergyCostBuy}{\ensuremath{\pi_{E}}}

\newcommand{\specificPowerCost}{\ensuremath{\pi_{P}}}
\newcommand{\energyCost}{\ensuremath{{C}_{E}}}
\newcommand{\powerCost}{\ensuremath{{C}_{P}}}
\newcommand{\totalCost}{\ensuremath{{C}}}

\newcommand{\timeIdx}{\ensuremath{k}}
\newcommand{\horizonLength}{\ensuremath{N}}
\newcommand{\inputVariable}{\ensuremath{u}}
\newcommand{\stateVariable}{\ensuremath{x}}
\newcommand{\gridUtilization}{\ensuremath{\gamma}}

\renewcommand{\thesection}{\arabic{section}}
\makeatletter
\def\@seccntformat#1{\csname iccas@#1@seccntformat\endcsname}
\def\iccas@section@seccntformat{\thesection.\quad}
\def\iccas@subsection@seccntformat{\thesubsection\quad}
\def\iccas@subsubsection@seccntformat{\thesubsubsection\quad}
\def\iccas@paragraph@seccntformat{\theparagraph\quad}
\makeatother

\begin{document}

\title{Optimal Electric Bus Depot Charging: Cost Savings, Grid Limits, and Robustness Trade-Offs}

\author{Fabio Widmer${}^{1*}$,
        Luca Pinter${}^{1}$,
        Mohammad Hossein Moradi${}^{1}$, and
        Christopher Harald Onder${}^{1}$}

\affils{${}^{1}$Institute for Dynamic Systems and Control (IDSC),
ETH Zurich, 8092 Zurich, Switzerland \\
{\small${}^{*}$Corresponding author (fawidmer@idsc.mavt.ethz.ch)}
}


\abstract{
Depot charging of electric bus fleets must minimize electricity costs, respect
grid limits, and remain feasible despite uncertain trip energy demand. While
cost-optimal charging is well studied, its value under different electricity
prices and grid connection capacities, as well as the economic cost of
robustness, remain poorly quantified.
We address these gaps with a convex robust formulation in which bounded demand
uncertainty is enforced through worst-case state-of-energy constraints. The
formulation is evaluated against charge-on-arrival using realistic service
schedules for four Swiss depots containing 7--35 buses.
In the depots studied, smaller depots achieve the greatest relative benefit from
optimization, with total electricity cost reductions exceeding 50\%, because
optimization mitigates charging peaks that strongly affect their costs.
For all depot sizes, the savings from optimization increase with electricity
price volatility.
Optimization can also lower the grid capacity required for feasible operation by
over 40\%, although tight limits reduce peak shaving potential. 
Protection against energy-demand deviations of 10\% increases total electricity
cost by less than 0.1\%.
The resulting charging power profiles exhibit interpretable price-threshold and
peak-shaping behavior, providing practical guidance for real-world
implementations.}


\maketitle


\section{Introduction}

\subsection{Motivation}

Battery-electric buses eliminate tailpipe emissions but make depot charging a
fleet-level coordination problem. Each bus must recover its trip energy before
departure despite time-varying electricity prices, demand charges, a shared grid
connection, and uncertain energy consumption. Charging on arrival can therefore
create costly peaks, while charging plans based only on nominal demand may be
infeasible.

These challenges motivate depot charging strategies that are simultaneously cost-optimal, robust against energy-demand uncertainty, and sufficiently interpretable for practical deployment.

\subsection{Related Work and Research Gap}

\subsubsection{Optimization-Based Charging}

A substantial body of literature addresses cost-optimal overnight depot charging for electric bus fleets. Studies formulate mixed-integer or quadratic programs that coordinate individual bus charging decisions, subject to charger-port limits, depot power caps, and \ac{SOE} constraints, with objectives ranging from energy-cost minimization to joint energy-cost and battery-degradation management~\cite{Verbrugge2022,Houbbadi2019QP}. A consistent finding is that coordinated or \enquote{smart} charging outperforms naive charge-on-arrival strategies by 7--15\% in energy costs across realistic case studies~\cite{Verbrugge2022,SmartDepot2025}. The inclusion of demand charges strengthens this case further: early charging-station planning work demonstrates that jointly optimizing the peak power level and the charging power profile yields additional bill reductions beyond \ac{TOU} arbitrage alone~\cite{Leou2017}, and depot models incorporating distributed energy resources confirm that demand-charge management amplifies the value of charging flexibility~\cite{Arif2020}. Further analyses show that load-shifting potential scales with the temporal volatility of the electricity price and \ce{CO2}-intensity profile~\cite{Rupp2020}.

Despite this body of work, existing depot studies typically report aggregate savings for a specific tariff structure and fixed infrastructure configuration. 
The first research question is therefore:
\emph{How do total electricity cost savings from optimization relative to charge-on-arrival vary with electricity price spread and grid connection capacity?}

\subsubsection{Uncertainty Treatment and Robust Optimization}

Trip energy consumption exhibits significant variability, with empirical studies reporting inter-trip differences on the same route mostly attributable to temperature and driving-style effects~\cite{Vepseleinen2018}. Approaches to handling this uncertainty include rolling-horizon re-optimization with updated forecasts and stochastic scheduling. Robust optimization with bounded uncertainty sets has been applied to electric bus problems at the \emph{planning} level, covering integrated charger deployment and battery sizing using box and distributionally robust energy-consumption models~\cite{Zhou2022,Momen2025}, and to \emph{en-route} charge scheduling via Bertsimas--Sim budgeted uncertainty counterparts~\cite{Zeng2024}. The only operational scheduling study that applies budgeted robust sets specifically to \ac{BEB} energy demand at the depot level addresses a station with \ac{V2G} and demand-response capabilities~\cite{Kang2025}.

No prior work was identified that quantifies the \enquote{price of robustness}, in the sense of Bertsimas and Sim~\cite{BertsimasSim2004}, has not been quantified in the context of overnight depot charging.
Hence, the second research question is:
\emph{What additional total electricity cost is incurred to protect the optimization solution against bounded trip energy-demand uncertainty?}

\subsection{Contributions}

The paper makes three main contributions:
\emph{First}, we present a convex robust optimization model for overnight \ac{BEB} depot charging.
\emph{Second}, we systematically compare the robust optimization solution with a charge-on-arrival baseline across a range of electricity price spreads and normalized grid connection capacities, identifying the operating regimes in which optimization delivers the greatest cost savings.
\emph{Third}, we present explicit cost--robustness trade-off curves that show how total electricity cost increases with energy-demand uncertainty, providing actionable guidance for practitioners selecting a robustness level under a combined energy and demand-charge tariff structure.

\section{Mathematical Model}
\label{sec:depot-charging-model}

\subsection{Depot Charging}

We consider a bus depot serving \Nbus{} buses, each with its own charger.
We use $\bus \in \busSet = \{1, 2, \ldots, \Nbus\}$ to enumerate all buses.
We use $\Pchgbus(\timeVar)$ to denote the power delivered to the charger of bus \bus{}.
The overall power drawn from the grid is
\begin{equation}
    \Pgrid(t)=\sum_{\bus\in\busSet}\Pchgbus(t).
    \label{eq:power-balance}
\end{equation}
We assume that the individual chargers and the grid connection are
unidirectional, i.e.,
\begin{align}
    0&\leq\Pchgbus(t)\leq\PchgbusLim, \label{eq:charger-limits}\\
    0&\leq\Pgrid(t)\leq\PgridUpperLim. \label{eq:grid-limit}
\end{align}
Each charger is rated at $\PchgbusLim=\qty{150}{kW}$.
\PgridUpperLim{} represents the maximum power that can be drawn from the grid.
Unless specified otherwise, the case studies assume that this limit is
high enough to allow simultaneous charging of all buses at full power.

The \ac{SOE} $\Ebus(t)$ of bus \bus{} evolves as
\begin{equation}
    \frac{\mathrm d}{\mathrm dt}\Ebus(t)=\Pibus(t)-\Pdemandbus(t), \label{eq:bus-ode}
\end{equation}
where $\Pibus(t)$ is battery charging power and $\Pdemandbus(t)$ is traction and
auxiliary demand. 
While bus \bus{} is in operation, its power demand depends on factors including
the driving mission, passenger load, and ambient conditions.
Because this paper focuses on depot charging optimization, we combine all
contributions to the bus power demand in $\Pdemandbus(t)$.
These contributions include propulsion power, driving losses, auxiliary
consumption, and battery losses.
Therefore, $\Pdemandbus(t)$ enters directly on the right-hand side of \cref{eq:bus-ode}.

We model the \ac{SOE} of each bus as the energy stored relative to its lowest admissible \ac{SOE}.
Its upper limit is represented by \EbusLim{}, which accounts for restrictions
imposed by the \ac{BMS}, for example to limit battery aging, and by the vehicle
operator, for example to maintain an emergency reserve:
\begin{equation}
    0 \leq \Ebus(t) \leq \EbusLim.
    \label{eq:bus-soe-limits}
\end{equation}

Charging losses are modeled by
\begin{equation}
    \Pibus(t)=\effbus\left(\Pchgbus(t)-\chgLossBus\Pchgbus^2(t)\right),
    \label{eq:bus-loss}
\end{equation}
with $\effbus=95\%$ and $\chgLossBus=\qty{3.51e-7}{\per\watt}$.

The known dispatch indicator $\inDepot(t)$ equals one when a bus is parked and
zero otherwise. Charging availability is therefore limited by
\begin{equation}
    0 \leq\Pchgbus(t)\leq\PchgbusLim\cdot \inDepot(t), \label{eq:bus-charge-in-depot-constraint}
\end{equation}
while the prescribed demand profile satisfies $\Pdemandbus(t)=0$ whenever $\inDepot(t)=1$.

\subsection{Energy-Demand Uncertainty}

The dispatch schedule is known, but the nominal energy-demand profile is considered
uncertain. For a relative uncertainty level $\demandUncertainty\geq0$, every
realization $\PdemandbusRealization(t)$ satisfies
\begin{equation}
 (1-\demandUncertainty)\cdot\Pdemandbus(t)
 \leq\PdemandbusRealization(t)
 \leq(1+\demandUncertainty)\cdot\Pdemandbus(t).
 \label{eq:demand-uncertainty-bounds}
\end{equation}
Hence, a worst-case power demand profile is given by $\PdemandbusWorst(t)=(1+\demandUncertainty)\cdot\Pdemandbus(t)$. 
A worst-case state driven by the same charging power,
\begin{align}
 \frac{\mathrm d}{\mathrm dt}\EbusWorst(t)
 &=\Pibus(t)-\PdemandbusWorst(t),\label{eq:bus-worst-case-ode}\\
 \EbusWorst(t)&\geq0,
\end{align}
guarantees the lower \ac{SOE} bound for every admissible realization. 
A separate upper-envelope trajectory for
low-demand realizations is unnecessary to guarantee feasibility, since charging power can always be reduced
when a bus approaches its upper \ac{SOE} limit.

\subsection{Electricity Cost}

The objective combines the energy cost induced by time-varying electricity
prices with a demand charge:
\begin{align}
 \totalCost&=\energyCost+\powerCost, \label{eq:grid-cost}\\
 \energyCost&=\int_{\tInit}^{\tFinal}
 \specificEnergyCostBuy(t)\cdot \Pgrid(t)\,\mathrm dt,\\
 \powerCost&=\specificPowerCost\cdot
 \max_{t\in[\tInit,\tFinal]}\Pgrid(t).
 \label{eq:power-cost}
\end{align}
The case studies use a monthly demand-charge rate of
\qty{12.1}{\$\per\kilo\watt}, normalized to the optimization horizon.

\section{Robust Charging Optimization}
\label{sec:robust-charging-optimization}

The goal is to determine the charging power of each bus over the planning
horizon such that the total electricity cost, comprising energy cost and the
demand charge, is minimized. The resulting charging power profile must satisfy
the depot power limits and nominal \ac{SOE} constraints while ensuring
feasibility under worst-case energy-demand assumptions.

\subsection{Convexifications}

To obtain a tractable convex optimization problem, we reformulate the
nonconvex battery-loss, depot-availability, and demand-charge constraints.

The non-affine equality constraint \cref{eq:bus-loss} that describes the battery model can be relaxed to obtain the convex constraint
\begin{equation}
    \Pibus(t) \leq \effbus \cdot \left( \Pchgbus(t) - \chgLossBus \cdot \Pchgbus^2 (t) \right).
    \label{eq:bus-loss-relaxed}
\end{equation}
This relaxation enlarges the feasible set relative to the original equality
constraint in \cref{eq:bus-loss}.
However, solutions with a strict inequality are suboptimal because they
correspond to additional energy losses.
Hence, we expect the optimal solution to satisfy \cref{eq:bus-loss} with equality, which we check a posteriori.

The binary constraint \cref{eq:bus-charge-in-depot-constraint} can be reformulated to
\begin{equation}
    \Pchgbus(t) \leq \PchgbusLim \cdot \inDepot(t).
\end{equation}

The nonconvex maximum term in \cref{eq:power-cost} can be reformulated as
follows:
\begin{align}
    \Pmax &\ge \Pgrid(t) \quad \forall\, t \in [\tInit, \tFinal], \label{eq:power-cost-relaxed-1} \\
    \powerCost &= \specificPowerCost \cdot \Pmax, \label{eq:power-cost-relaxed-2}
\end{align}
Similarly to the battery-loss relaxation, the optimal solution is expected to satisfy 
\cref{eq:power-cost-relaxed-1} tightly, which we check a posteriori.

\subsection{Discrete-Time Formulation}

After the above reformulations, the continuous-time \ac{OCP} is transcribed
into a finite-dimensional convex program using multiple shooting and forward
Euler integration with a uniform discretization time step.
We use \timeIdx{} for the discrete time index.

\newcommand{\transposed}{\mathsf{T}} 

For notational simplicity, we first introduce the input variable vector $\inputVariable_{\timeIdx{}}$ and the state variable vector $\stateVariable_{\timeIdx{}}$:
\begin{align}
\inputVariable_{\timeIdx{}} &= \left[
\Pchgbus{}[\timeIdx], \Pibus{}[\timeIdx], \Pgrid{}[\timeIdx]
\right]^{\transposed} \notag \\
& \hspace{3cm}\text{for }\timeIdx \in \{0, \ldots, \horizonLength-1 \},  \\
\stateVariable_{\timeIdx{}} &= \left[
\Ebus{}[\timeIdx], \EbusWorst{}[\timeIdx]
\right]^{\transposed} \quad \text{for } \timeIdx \in \{0, \ldots, \horizonLength \},
\end{align}
where $\Pchgbus{}[\timeIdx]$, $\Pibus{}[\timeIdx]$,
$\Ebus{}[\timeIdx]$, and $\EbusWorst{}[\timeIdx]$ each contain one entry for
every $\bus \in \busSet$.

Using these variables, the finite-dimensional convex program is formulated as follows:

\newcommand{\forallk}{\; \forall\,\timeIdx}
\newcommand{\forallu}{\; \forall\,\bus}
\newcommand{\forallkandu}{\; \forall\,\timeIdx ,\,\bus}

\begin{subequations}\label{eq:optproblem}%
\small
\begin{align}
     \multicolumn{2}{l}{$\displaystyle
    \underset{\footnotesize \begin{matrix}
    \inputVariable_0, \ldots, \inputVariable_{\horizonLength-1} \\
    \stateVariable_0, \ldots, \stateVariable_{\horizonLength} \\
    \Pmax
    \end{matrix}
    }{\text{minimize}} \; \specificPowerCost \,\Pmax
    + \sum_{\timeIdx \in \{0, \ldots, \horizonLength-1\}}
    \specificEnergyCostBuy{}[\timeIdx]\, \Pgrid{}[\timeIdx]\, \timeStep
    $} \\
    \text{s.t.\;\;}
    & \Ebus{}[\timeIdx+1] = \Ebus{}[\timeIdx] + \timeStep \cdot (\Pibus{}[\timeIdx] - \Pdemandbus{}[\timeIdx]) \forallkandu \\
    & \EbusWorst{}[\timeIdx+1] = \EbusWorst{}[\timeIdx] + \timeStep \cdot \left(\Pibus{}[\timeIdx] - \PdemandbusWorst{}[\timeIdx]\right) \forallkandu \\
    & \Pibus{}[\timeIdx] \leq \effbus \cdot \left( \Pchgbus{}[\timeIdx] - \chgLossBus \cdot \Pchgbus^2 [\timeIdx] \right) \forallkandu\\
    & 0 \leq \Ebus{}[\timeIdx] \leq \EbusLim \forallkandu \\
    & 0 \leq \EbusWorst{}[\timeIdx] \forallkandu \label{eq:optproblem:robustness} \\
    & \Pgrid{}[\timeIdx] = \sum_{\bus \in \busSet} \Pchgbus{}[\timeIdx] \forallk \label{prb:balance}\\
    & 0 \leq \Pgrid{}[\timeIdx] \leq \PgridUpperLim \forallk \\
    & 0 \leq \Pchgbus{}[\timeIdx] \leq \PchgbusLim\cdot \inDepot{}[\timeIdx] \forallkandu \\
    & \Pmax \geq \Pgrid{}[\timeIdx] \forallk\\
    & \EbusWorst{}[0] = \Ebus{}[0] \forallu \label{eq:optproblem:wcinit} \\
    & \Ebus{}[\horizonLength] \ge \Ebus{}[0] \forallu \label{eq:optproblem:chargesustain}
\end{align}
\end{subequations}

The optimization horizon is embedded in a periodic operating cycle, so the initial \ac{SOE} is treated as a cyclic variable rather than a measured state.
The terminal constraint \cref{eq:optproblem:chargesustain} prevents the
optimizer from reducing cost by depleting the fleet at the horizon boundary.
This condition is appropriate when the dispatch and demand profiles are approximately periodic.
For public transport applications, the horizon should therefore
span an integer multiple of 24 hours. The present study uses a 24-hour horizon.

\section{Case Studies and Results}
\label{sec:case-studies-results}

\subsection{Case Study Setup}
\label{sec:case-study-setup}

Four realistic Swiss depot scenarios are derived from public \ac{GTFS}
schedules and elevation-based energy estimates. They range from 7 to 35 buses
(\cref{tab:depots}).

\begin{figure}
    \centering
    \includegraphics[width=\linewidth,
    trim=0 6mm 0 6mm,
    clip]{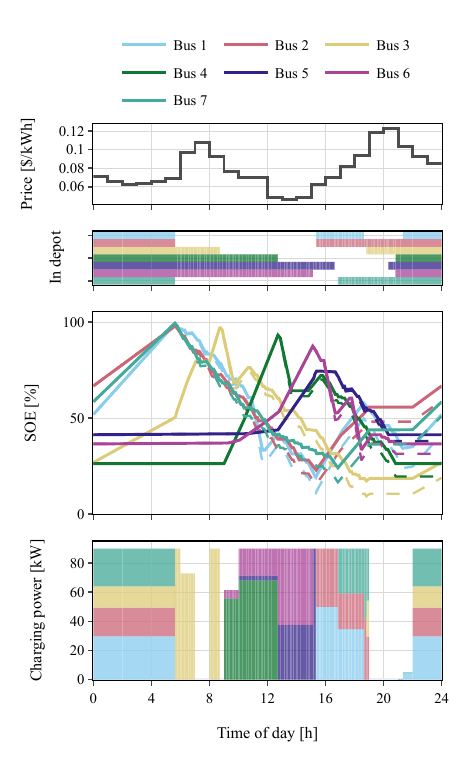}
    \caption{Optimized charging power profile for the Glarus depot with seven buses.
    The panels show the electricity price profile, each bus's depot availability, the optimized \ac{SOE} trajectories with dashed worst-case trajectories, and the charging power profiles.
    }
    \label{fig:opt-strategy}
\end{figure}

For the electricity price, a historical day-ahead profile
\cite{spot_price} defines the medium-volatility reference case (see the top
subplot in \cref{fig:opt-strategy}).
Flat, low-, and high-volatility profiles are
obtained by scaling deviations from its daily mean by factors 0, 0.5, and 1.5,
respectively. Thus, all profiles have the same mean electricity price.

\begin{table}[b]
\setlength{\extrarowheight}{0.75ex}
\caption{Overview of the four depots used in the case studies.}
\label{tab:depots}
\centering
\begin{tabularx}{\linewidth}{Xrrr}
Depot & Buses & Bus-hours & Energy [MWh] \\ \hline
Glarus & 7 & 61.0 & 1.62 \\
Einsiedeln & 14 & 133.8 & 4.53 \\
Zufikon & 21 & 250.4 & 7.77 \\
Bern & 35 & 372.4 & 13.90 \\
\end{tabularx}
\end{table}

\subsection{Strategy Comparison}
\label{sec:strategy-comparison}

The charge-on-arrival strategy serves as the heuristic baseline in this study. Under 
this policy, each bus begins charging at its maximum admissible power immediately upon 
returning to the depot and continues until the battery is fully replenished. 
Grid connection limits are enforced by allocating the available capacity as evenly as possible among active chargers, without exceeding each bus's requested charging power.
For example, requests of $100$, $100$, and \qty{50}{kW} under a \qty{200}{kW} grid limit result in allocations of $75$, $75$, and \qty{50}{kW}.

\begin{figure}
    \centering
    \includegraphics[width=\linewidth,
    trim=0 4mm 0 6mm,
    clip]{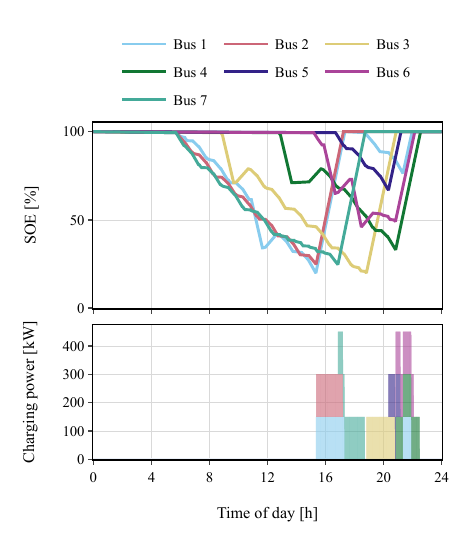}
    \caption{Charge-on-arrival strategy for the Glarus depot, for the same scenario as \cref{fig:opt-strategy}.}
    \label{fig:coa-strategy}
\end{figure}

In the charge-on-arrival case (\cref{fig:coa-strategy}), each bus initiates charging 
immediately upon returning to the depot, resulting in highly concentrated and largely 
uncontrolled demand peaks reaching approximately \qty{440}{kW} in the late afternoon and evening 
hours. By contrast, the optimized strategy uses larger portions of each bus's
in-depot availability window to redistribute the charging load over time and
reduce the fleet's peak charging power. When a higher power level is required
to maintain feasibility, the optimizer concentrates the corresponding energy
throughput in the lowest-price intervals, yielding a cost-efficient charging
power profile with a lower grid peak.

The optimized strategy shifts charging toward low-price periods, mainly early morning and midday,
while avoiding the late-morning and evening price peaks.
In contrast, charge-on-arrival partly overlaps with the evening peak and therefore incurs higher energy costs.

The practical benefits of the optimized charging strategy are reflected in the
key performance indicators. Compared with the charge-on-arrival baseline, the optimized
charging power profile reduces peak grid power, and therefore the demand
charge, by 80\%. It also reduces energy cost by 26\%, yielding a 54\%
reduction in total electricity cost.

Beyond the reductions in cost and peak power, the optimized profile exhibits a clear structure:
above an implicit price threshold, chargers remain idle;
below it, charging fills an approximately rectangular power-time profile.
Minor deviations from a perfectly rectangular profile arise due to quadratic 
conduction losses in the charger and battery systems, which slightly penalize higher 
power levels.
This interpretable solution structure facilitates validation and can inform
practical heuristic charging strategies for fleet operators.


\subsection{Cost of Robustness}
\label{sec:cost-of-robustness}

The robustness constraint \cref{eq:optproblem:robustness} requires feasibility
under worst-case energy-consumption realizations by tightening the feasible region,
which can increase total electricity cost relative to the nominal solution.

\begin{figure}
    \centering
    \includegraphics[width=\linewidth]{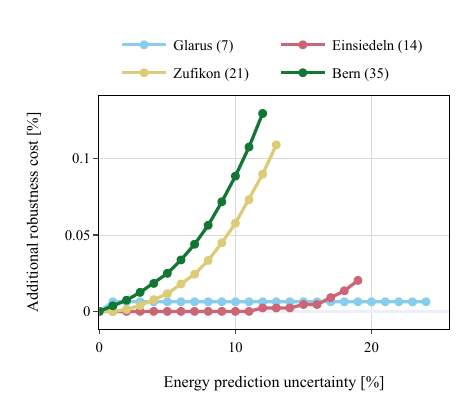}
    \caption{Trade-off between the assumed worst-case energy-demand uncertainty and the resulting increase in total electricity cost.}
    \label{fig:cost-of-robustness}
\end{figure}

\Cref{fig:cost-of-robustness} illustrates this cost--robustness trade-off across 
the four depots. Two key observations follow. First, larger depots tend to be more 
sensitive to energy-demand uncertainty: the Bern depot (35~buses) incurs
robustness costs more rapidly than the Glarus depot (7~buses) for the same uncertainty
level.
Second, the absolute cost increase remains small across all scenarios,
indicating that worst-case feasibility requires only minor adjustments to the
nominal charging power profile.
This behavior is consistent with the robustness constraint being inactive for
many missions, as illustrated for instance in \cref{fig:opt-strategy}.

\subsection{Sensitivity of Price Volatility and Depot Size}
\label{sec:price-volatility-depot-size}

\begin{figure}
    \centering
    \includegraphics[width=\linewidth]{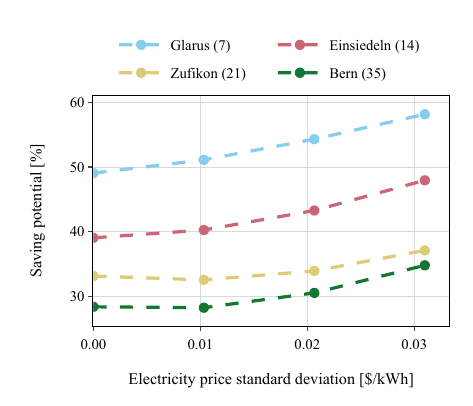}
    \caption{Influence of electricity price volatility on the total electricity cost savings achieved through charging optimization.}
    \label{fig:sensitivity-price-volatility}
\end{figure}

To assess the influence of electricity price dynamics on the achievable savings, four 
price profiles are considered. 
The results, presented in \cref{fig:sensitivity-price-volatility}, reveal two consistent 
trends. First, regardless of depot size, higher price volatility translates directly into 
greater savings: as the temporal spread of electricity prices widens, the
optimizer gains more opportunity to shift charging load toward low-price periods, yielding 
reductions in total electricity cost ranging from approximately 25\% under a
flat electricity price profile to
nearly 60\% under the most volatile profile. 
Second, smaller depots consistently exhibit greater savings potential across the volatility range.
With fewer buses, aggregate charging demand is more sensitive to randomly occurring price peaks, whereas this variability is partially smoothed by fleet aggregation in larger depots.
Optimization therefore achieves larger relative savings in smaller depots by reducing peak power demand, as indicated in \cref{fig:opt-strategy,fig:coa-strategy}.

\subsection{Value of Optimization Under Grid Constraints}
\label{sec:grid-constraints}

To quantify the influence of the available grid connection on charging performance, we define the 
grid utilization factor as
\begin{equation}
    \gridUtilization = \frac{\PgridUpperLim}{\sum_{\bus \in \busSet} \PchgbusLim}.
\end{equation}
A value of $\gridUtilization = 1$ therefore corresponds to an 
unconstrained scenario in which the grid connection is large enough to charge all buses 
simultaneously at full power.
This assumption applies to all preceding case studies.
Smaller values of \gridUtilization{} represent increasingly stringent grid 
limitations.

\begin{table}
    \centering
    \caption{Minimum feasible grid utilization factor $\gridUtilization$ per depot for the 
    charge-on-arrival and optimized charging strategies, along with the relative reduction achieved through optimization.}
    \label{tab:min-grid-utilization}
\begin{tabularx}{\linewidth}{Xrrr}\small
Depot & Charge-on-arrival & Optimal & Reduction [\%] \\ \hline
Glarus & 0.095 & 0.095 & 0.0 \\
Einsiedeln & 0.143 & 0.119 & 17 \\
Zufikon & 0.222 & 0.127 & 43 \\
Bern & 0.200 & 0.124 & 38 \\
\end{tabularx}
\end{table}

\begin{figure}
    \centering
    \includegraphics[width=0.95\linewidth,
    trim=0 1mm 0 6mm,
    clip]{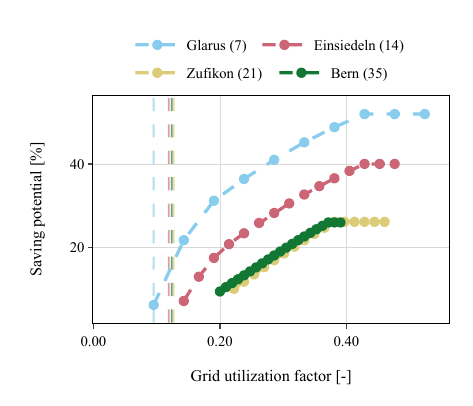}
    \caption{Influence of the grid utilization factor (a lower factor represents a more stringent grid limitation) on the total electricity cost savings achieved through charging optimization.
    The vertical dashed lines indicate the minimum grid utilization factor for which the charging optimization remains feasible.}
    \label{fig:grid-lim-sensitivity}
\end{figure}

\Cref{fig:grid-lim-sensitivity} shows total electricity cost savings as a
function of \gridUtilization{}
for the four depots considered, with the vertical dashed lines indicating the minimum 
$\gridUtilization$ at which the charging optimization remains feasible.
\Cref{tab:min-grid-utilization} provides a comprehensive overview of these minimum 
feasible values across all depots, comparing the charge-on-arrival and optimized 
strategies. Two key observations emerge. First, the optimized strategy substantially 
reduces the minimum grid utilization factor required for feasibility, with reductions 
ranging over 40\%. 
This reduction is particularly relevant for larger depots: optimized charging
can enable fleet electrification with only 60\% the grid connection capacity that would otherwise be
required under a charge-on-arrival policy. Second, for smaller depots, the minimum 
feasible $\gridUtilization$ is already inherently low even without optimization (e.g., 
$\gridUtilization = 0.095$ for Glarus), leaving comparatively less room for further reduction. 

A third effect visible in \cref{fig:grid-lim-sensitivity} is that higher values of 
$\gridUtilization$ are associated with greater total electricity cost savings.
Intuitively, with a less restrictive grid
connection, the charge-on-arrival strategy tends to produce high power peaks, whereas the
optimizer can redistribute charging across time and shift consumption toward lower-price
periods. Under stringent grid limitations, charging optimization instead primarily ensures operational
feasibility with a smaller grid connection, potentially reducing infrastructure investment costs
and improving the economic case for fleet electrification.

\section{Conclusion}
\label{sec:conclusion-outlook}

Across four realistic depots, optimized charging reduced total electricity costs by
25--60\% as price volatility increased and lowered the grid capacity required
for feasible operation by over 40\%. Tight grid limits reduce cost-shifting
flexibility, but make optimization more important for infrastructure
feasibility. Protection against energy-demand deviations of 10\% increased
total electricity cost by less than 0.1\%. The resulting price-threshold and peak-shaping
behavior are also operationally interpretable.

Future work will address closed-loop re-optimization for uncertain arrivals
and consumption, followed by extensions to photovoltaic generation,
stationary storage, and building demand.

The observed interaction between grid constraints, price volatility, and robustness requirements also motivates a broader depot-sizing question: how should fleet size and grid connection capacity be jointly selected, and which operational and economic factors govern their optimal ratio? Addressing this design problem requires extending the present operational optimization framework to include infrastructure-sizing decisions.

\section*{Acknowledgement}

The authors gratefully acknowledge PostAuto Schweiz AG for valuable discussions and for providing domain expertise and practical perspectives on real-world fleet operations, in the context of the e-Sandbox initiative under MobilityLab Sion.
This work was financially supported by the Zurich Information Security and Privacy Center (ZISC) at ETH Zurich.

\bibliographystyle{IEEEtran}
\bibliography{references}

\end{document}